\documentclass{emulateapj}

\shorttitle{GRB central engine}
\shortauthors{Janiuk et al.}

\begin{document}

\title{Accretion and outflow from a magnetized, 
neutrino cooled torus around the gamma ray burst central engine}

\author{Agnieszka Janiuk$^{1}$, Patryk Mioduszewski$^{1}$}
\affil{$^{1}$
Center for Theoretical Physics, Polish Academy of Sciences, 
Al. Lotnikow 32/46, 
02-668 Warsaw, Poland}
\email{agnes@cft.edu.pl}
\and
\author{Monika Moscibrodzka$^{2}$}
\affil{$^{2}$ Department of Physics, University of Nevada Las Vegas, 
4505 South Maryland Parkway, Las Vegas, NV 89154, USA
}

\begin{abstract}
We calculate the structure and short-term evolution of a gamma ray burst
central engine in the form of a turbulent torus accreting onto a stellar mass
black hole. Our models apply to the short gamma ray burst events, in which a
remnant torus forms after the neutron star-black hole or a double neutron star
merger and is subsequently accreted.  We study the 2-dimensional, relativistic
models and concentrate on the effects of black hole and flow parameters as
well as the neutrino cooling.  We compare the resulting structure and neutrino
emission to the results of our previous 1-dimensional simulations. We find
that the neutrino cooled torus launches a powerful mass outflow, which
contributes to the total neutrino luminosity and mass loss from the
system. The neutrino luminosity may exceed the Blandford-Znajek luminosity of
the polar jets and the subsequent annihillation of neutrino-antineutrino pairs
will provide an additional source of power to the GRB emission.
\end{abstract}

\section{Introduction}

Gamma Ray Bursts (GRB), known for about forty years \citep{klebes73} are
extremely energetic transient events, visible from the most distant parts of
the Universe. They last from a fraction of a second up to a few hundreds of
seconds and are isotropic, non-recurrent sources of gamma ray radiation (10
keV - 20 MeV). Short gamma ray bursts were distinguished in the KONUS data by
\citep{mazets81} and further two distinct classes of events, long and short,
were found by \citep{Kouvel93}.  The energetics of these events points to a
cosmic explosion as a source of the burst, associated with the compact objects
such as black holes and neutron stars.  The short timescales and high Lorentz
factors of the gamma ray emitting jets are most likely produced in the process
of accretion of rotating gas on the hyper-Eddington rates that proceeds onto a
newly born stellar mass black hole. The key properties of such a scenario are
therefore deep gravitational potential of the black hole and significant
amount of the angular momentum that supports the rotating torus.

Accretion of magnetized torus onto a black hole with a range of spin 
parameters was studied by \citet{devillers, McKinneyGammie2004} and 
applied to the long gamma ray bursts \citep{nagataki}. The relativistic 
simulations of accretion flows 
with an ideal gas equation of state were studied e.g., 
by \citet{HawleyKrolik2006} and \citet{McKinneyBlandford2009} and recently more 
sophisticated models with a realistic EOS were proposed by 
\citet{bkomis08, bkomis10} and \citet{barkov08, bbaushev2011}.

This central
engine gives rise to the most powerful jets (see e.g. the reviews by
\citet{Zhang04, Piran05, Gehlers09, Metzger}, \citet{Gehlers09},
\citet{Metzger}).  Despite the existence of still unsolved problems, such as
the composition of the outflows, the emission mechanisms creating the gamma
rays, or the form of energy that dominates the jet (i.e. kinetic or Poynting
flux), the jets themselves are believed to be powered by accretion and
rotation of the central black hole. In this process, the strong large-scale
magnetic fields play a key role in transporting the energy to the jets
\citep{McKinney06,Tchekov08,Dexter2012}.
 
In addition to the magneto-rotational mechanism of energy extraction, the
annihilation of neutrino-antineutrino pairs, emitted from the accreting
torus, may provide some energy reservoir available in the polar regions to
support jets. The neutrinos are produced in central engines of both short and
long GRBs, the latter being modeled in the frame of the collapsing massive
star scenario \citep{woosley93, paczynski98}.  The recent numerical
simulations of the 'hypernovae' aimed to capture the effects of both MHD and
neutrino transport in the supernova explosion modeling \citep{burrows07},
using a flux-limited neutrino diffusion scheme in the Newtonian dynamics.  The
general relativistic simulations by \citet{Shibata2007} on the other hand,
consider the neutrino cooling of the accreting torus around the black hole and
capture the neutrino-trapping effect in a qualitative way.

In this work, we study the central engine, composed of a stellar mass,
rotating black hole and accreting torus that has formed from the remnant
matter at the base of the GRB jet. We start from an axially symmetric,
configuration of matter filling the equipotential surfaces around a Kerr black
hole \citep{Moncrief, Abramowicz}, assuming an initial poloidal magnetic
field.  The MHD turbulence amplifies the field and leads to the transport of
angular momentum within the torus.  In the dynamical calculations, we
use a realistic equation of state while we
account for the neutrino cooling \citep{yuan, janiuk}.  We study the evolution
and physical properties of such an engine, its neutrino luminosity and
production of a wind and outflow from the polar regions.  Our calculations are
2-D and relativistic, therefore this work is a generalization of the model
presented in \citet{janiuk, janiuk2010}, where a simpler steady-state, 1
dimensional model of a torus around a rotating black hole was analyzed, using
approximate correction factors to the pseudo-Newtonian potential that allowed
to mimic Kerr metric. The microphysics however is currently 
described using the EOS from that work
and neutrino cooling is incorporated into the HARM scheme via the 
cooling function.
The total pressure invoked to compute the cooling
is contributed by the free and degenerate nuclei,
electron-positron pairs, helium, radiation and partially trapped neutrinos. This allows us 
to compute the optical depths for neutrino absorption and scattering and the 
neutrino emissivities in the optically thin or thick plasma.

The article is organized as follows. In \S~\ref{sec:model}, 
we describe our model, the
initial conditions, the dynamical evolution of the system and the assumed
chemical composition as well as the processes responsible for energy losses
via neutrino cooling. In \S~\ref{sec:results}, we present the results, describing the
effects of (i) black hole mass (ii) its spin (iii) torus mass, and (iv)
magnetic field strength. We also discuss the effect of neutrino cooling on the
torus structure, in comparison with the reference model with such cooling
neglected. Finally, we compare our results with the 1-D simulations of the
vertically averaged torus, emphasizing the effects of 2-dimensional
computations.  We discuss the results in \S~\ref{sec:diss}.

\section{Model of the hyperaccreting disk}\label{sec:model}	
The model computations are based on the axisymmetric, general relativistic
MHD code {\it HARM-2D}, described by \citet{gammie} and \citet{noble}. The
nuclear equation of state is discussed in detail in \citet{janiuk}. The goal
of our calculations is to investigate the overall structure of a magnetized,
turbulent accretion disk in which nuclear reactions take place and the gas
looses energy via neutrino cooling, and in particular to expand our previous
1-dimensional models based on $\alpha$ viscosity, to the case of 2-D GRMHD.

\subsection{Initial conditions and dynamical model}
We start the numerical calculations from the equilibrium model of a
thick torus around a spinning black hole as introduced by
\citet{Moncrief} and \citet{Abramowicz}. The parameters of the model
are the central black hole mass, $M_{\rm BH}=3-10 M_{\odot}$, the
dimensionless spin of the black hole, $a=0.8-0.98$, and the total
mass of the surrounding gas, $M_{\rm torus}=0.1-2.5 M_{\odot}$ (see
Table~\ref{table:models} for the list of models). We seed the torus
with a poloidal magnetic field (magnetic field lines follow the
constant density surfaces); the strength of the initial magnetic field
is normalized by the gas to magnetic pressure ratio at the pressure
maximum of the initial structure of the disk
($\beta=P_{\rm gas}/P_{\rm mag}=5-100$). In the dynamical calculations, we 
use $P=(\gamma-1)u$ equation
of state with the adiabatic index $\gamma=4/3$. To follow the
evolution of the gas dynamics near a black hole we use a numerical MHD
code {\it HARM-2D}. The numerical code is designed to solve
magnetohydrodynamic equations in the stationary metric around a black
hole. 

In this work, we modify the MHD code to account for the chemical
composition of the nuclear matter accreting onto black hole in the GRB
environment (described in more detail in \S~\ref{sec:nuclear}). At
each time moment of the simulation we calculate the gas nuclear
composition assuming the balance of nuclear equilibrium
reactions. This gives us expected neutrino cooling rates which we
incorporate into the code. After each time step of the dynamical
evolution the total internal energy of gas is reduced by $Q_{\nu}
\Delta t$ factor using an explicit method with $n$-sub-cycles.
The procedure for calculating the neutrino cooling takes into account 
the change of the gas internal energy in the comoving frame, which is a 
correct relativistic approach. We do not account for the neutrino transfer 
though, and the effects like the gravitational redshift are neglected.

Our models have numerical resolution of the grid 256x256 points
in $r$ and $\theta$ directions (see also Sect. \ref{sect:resol}). 
The grid is logarithmic in radius and
condensed in polar direction towards the equatorial plane, as in Gammie et
al. (2003).

\subsection{Chemical composition and neutrino cooling}\label{sec:nuclear}

We assume that the neutrino emitting plasma consists of protons,
electron-positron pairs, neutrons and helium nuclei. The gas is in beta
equilibrium, so that the ratio of protons to neutrons satisfies the balance
between forward and backward nuclear reactions.

Neutrinos are formed in the URCA process, electron positron pair annihilation,
nucleon - nucleon bremsstrahlung, plasmon decay. These reactions are:
\begin{eqnarray}
\label{eq:ur1}
p + e^{-} \to n + \nu_{\rm e} \nonumber \\
\label{eq:ur2}
n + e^{+} \to p + \bar\nu_{\rm e} \nonumber \\
\label{eq:ur3}
n \to p + e^{-} + \bar\nu_{\rm e} \nonumber \\
\end{eqnarray}
 and
\begin{equation}
\tilde \gamma \to \nu_{\rm e}+\bar\nu_{\rm e}
\label{eq:plasmon}
\end{equation}
and
\begin{equation}
e^{-}+e^{+}\to \nu_{\rm i}+\bar\nu_{\rm i}
\label{eq:annihil}
\end{equation}
and
\begin{equation}
n+n \to n+n+\nu_{\rm i}+\bar\nu_{\rm i}.
\label{eq:brems}
\end{equation}

For a given temperature and density, the neutrino cooling rate is
calculated from the balance between the above reactions, supplemented with
the conditions of the conservation of the baryon number and charge neutrality
(Yuan 2005; see also Kohri \& Mineshige 2002, Chen \& Beloborodov 2007, 
Janiuk al. 2007).

We assume that the cooling proceeds via electron, muon and
tau neutrinos in the plasma opaque to their absorption and scattering.
The URCA process and plasmon decay produce the electron neutrinos only, while the other processes 
produce neutrinos of all flavors. The emissivities of these processes are
\begin{equation}
q_{\rm brems} = 3.35\times 10^{27} \rho_{10}^{2} T_{11}^{5.5}\;, 
\end{equation}
\begin{equation}
q_{\rm plasmon} = 1.5\times 10^{32} T_{11}^{9} \gamma_{p}^{6} e^{-\gamma_{p}}
(1+\gamma_{p})\left(2+{\gamma_{p}^{2} \over 1+\gamma_{p}}\right)\;, 
\end{equation}
\begin{equation}
q_{\rm e^{+}e^{-}} = q_{\nu_{\rm e}}+q_{\nu_{\mu}}+q_{\nu_{\tau}}
\end{equation}
and
\begin{equation}
q_{\rm urca} = q_{p+e^{-}\to n+\nu_{\rm e}} + q_{\rm n+e^{+}\to p+\bar\nu_{\rm e}} + q_{n\to p+e^{-}+\bar\nu_{\rm e}}\;.
\end{equation}
the two latter being iterated numerically (the full set of Equations
is given in the Appendix of Janiuk et al. 2007).
Here $\rho_{10}$ is the baryon density in the units of 10$^{10}$ g/cm$^{3}$ and
$T_{11}$ is temperature in the units of 10$^{11}$ K. The emissivities are
given in the units of [erg cm$^{-3}$s$^{-1}$].
We neglect here the term of neutrino cooling by photodissociation of 
helium nuclei, since at the temperatures and densities obtained in the 
presented models, this term will be practically equal to zero.

The plasma can be opaque to neutrinos, so we use the optical depths,
given by the equations derived in Di Matteo et al. 2002 :
\begin{equation}
\tau_{\rm a,\nu_{i}} = {H \over 4 {7 \over 8}\sigma T^{4}} q_{\rm a, \nu_{i}},
\end{equation}
where absorption of the electron neutrinos is determined by
\begin{equation}
q_{\rm a, \nu_{e}} = q_{\nu_{e}}^{\rm pair} + q_{\rm urca} + q_{\rm plasm} + {1\over 3} q_{\rm brems},
\end{equation}
and for the muon and tau neutrinos is given by
\begin{equation}
q_{\rm a, \nu_{\mu,\tau}} = q_{\nu}^{\rm pair} + {1\over 3} q_{\rm brems}\;.
\end{equation}
We also account for the neutrino scattering and the scattering optical depth is given by:
\begin{eqnarray}
\tau_{\rm s} &=& \tau_{\rm s,p}+\tau_{\rm s,n}  \\
&=& 24.28\times 10^{-5}
\nonumber \left[\left({kT\over m_{\rm e}c^{2}}\right)^{2} H \left(C_{\rm s,p} n_{\rm p} + C_{\rm s,n} n_{\rm n}\right)\right]
\end{eqnarray}
where $C_{\rm s,p}=(4(C_{V}-1)^{2}+5\alpha^{2})/24$,  $C_{\rm s,n}=(1+5\alpha^{2})/24$,
$C_{\rm V}=1/2+2 \sin^{2}\theta_{\rm C}$, with $\alpha=1.25$ and $\sin^{2}\theta_{\rm C}=0.23$ \citep{yuan, reddy}.

The neutrino cooling rate is finally given by
\begin{equation}
Q^{-}_{\nu} = { {7 \over 8} \sigma T^{4} \over 
{3 \over 4}} \sum_{i=e,\mu} { 1 \over {\tau_{\rm a, \nu_{i}} + \tau_{\rm s} \over 2} 
+ {1 \over \sqrt 3} + 
{1 \over 3\tau_{\rm a, \nu_{i}}}} \times {1 \over H}\; ~[{\rm erg ~s^{-1} ~cm^{-3}]}
\label{eq:qnuthick}
\end{equation}
and the neutrino luminosity emitted by the plasma is
\begin{equation}
L_{\nu} = \int{Q^{-}_{\nu} dV} ~[{\rm erg ~s^{-1}]}.
\label{eq:lumin}
\end{equation}
where $dV$ is the unit volume in the Kerr geometry.

The optical depths for absorption and scattering are calculated approximately
by assuming the disk vertical thickness equal to the pressure scale-height,
$H=c_{\rm s}/\Omega_{\rm K}$, where $c_{\rm s}$ is the speed of sound and
$\Omega_{\rm K} = {c^{3} \over GM_{\rm BH}} (a+r^{3/2})^{-1}$ is the Keplerian frequency
(see e.g. \citealt{Lopez09}). 
The resulting thickness
is roughly 
proportional to a fraction of the disk radius and the typical ratios are
$H/r \sim 0.3-0.5$.

We do not account for the neutrino heating in the jets via the annihilation
process, 
because of large uncertainties in the internal energy computations in the jet.

The neutrino cooling is limited to the torus and wind only, via
the density and temperature ranges for which the cooling is operating
($10^{6}-10^{13}$ g cm$^{-3}$ and $10^{7}-10^{12}$ K, respectively).
Therefore  the jets are not shown in the neutrino cooling maps.

\section{Results}\label{sec:results}

\subsection{Effect of the BH parameters and torus mass on the $\dot{M}$
and neutrino luminosity}

\begin{figure}
\includegraphics[width=7cm]{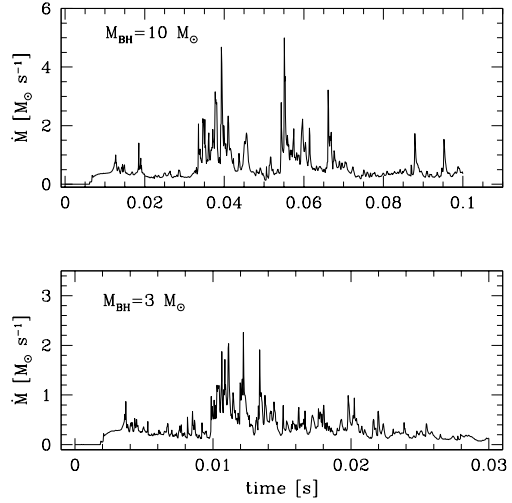}
\caption{Mass accretion rate onto black hole as a function of time.
  The black hole mass is $M_{\rm BH}=3
  M_{\odot}$ and torus initial mass is $M_{\rm d} \sim 0.1 M_{\odot}$ (bottom panel), or
  $M_{\rm BH}=10 M_{\odot}$ and $M_{\rm d} \sim 1.0 M_{\odot}$ (top panel). 
  The black hole spin is $a=0.98$.}
\label{fig:accrate}
\end{figure}

We studied the models with the black hole mass of $M_{\rm BH}=3 M_{\odot}$ or
$M_{\rm BH}=10 M_{\odot}$, and the torus mass was assumed equal to about 0.1,
0.3, 0.7, 1.0 or 2.6 $M_{\odot}$ (Table~\ref{table:models}).
In Figure~\ref{fig:accrate}, we show the time evolution of the mass accretion
rate onto black hole, for the two values of torus and black hole mass. 
The average accretion rate onto black hole is not changing much
with the black hole spin and is about 0.3-1.0 $M_{\odot}$ s$^{-1}$ 
for most SBH models. The accretion rate
for the first 2-3 milliseconds is very small, and then grows to about 0.2-0.5
$ M_{\odot}$ s$^{-1}$ and starts varying. During such flares, it exceeds
momentarily 2-5 $ M_{\odot}$ s$^{-1}$. These flares are however very short in
duration. The mean accretion rate in our models does not exceed 1 $M_{\odot}$
s$^{-1}$.

The magnitude of the flares depends on the black hole spin, and largest is
for $a=0.8$ in the small disk models (SBH).
The amplitude of flares is by a factor of $\sim 2-3$ larger for the black hole
mass of 10 $M_{\odot}$ (LBH). 
In the LBH models, the case with $a=0.9$ shows higher flares at the early
evolution, while the $a=0.8$ model is flaring in the late times.  After the
time of about $t=3000 M$, the accretion rate decreases, the flaring ceases and
a rather stable value below $\dot{M} \lesssim 0.3 M_{\odot}$s$^{-1}$ is reached. 
The
late time activity ceases because of the 
decay of magnetic turbulence
characteristic for axisymmetric models.

\begin{figure*}
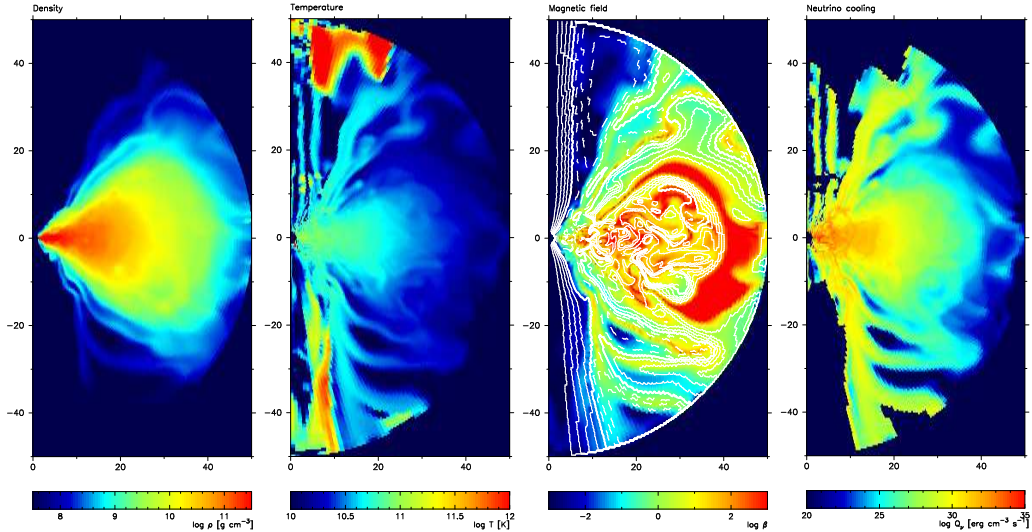

\centering
\includegraphics[width=7cm,angle=270]{Fig2_density_M3_v124.ps}
\includegraphics[width=7cm,angle=270]{Fig2_temperature_M3_v124.ps}
\includegraphics[width=7cm,angle=270]{Fig2_beta_M3_v124.ps}
\includegraphics[width=7cm,angle=270]{Fig2_neu_M3_v124.ps}
\caption{2-D model: Structure of accretion disk in model with neutrino cooling
  taken into account in the dynamical evolution.  The maps show: (i) density,
  (ii) temperature of the plasma, (iii) ratio of gas to magnetic pressure,
  with field lines topology, and (iv) the effective neutrino cooling $Q_{\nu}$
  (from left to right).  Parameters: black hole mass $M = 3M_{\odot}$, spin
  $a=0.98$, initial magnetic field normalization $\beta=50$, and initial disk
  mass $M_{\rm disk}= 0.1 M_{\odot}$. The snapshot is at t=0.03 s since the
  formation of the black hole.}\label{fig2}
\end{figure*}

\begin{figure*}
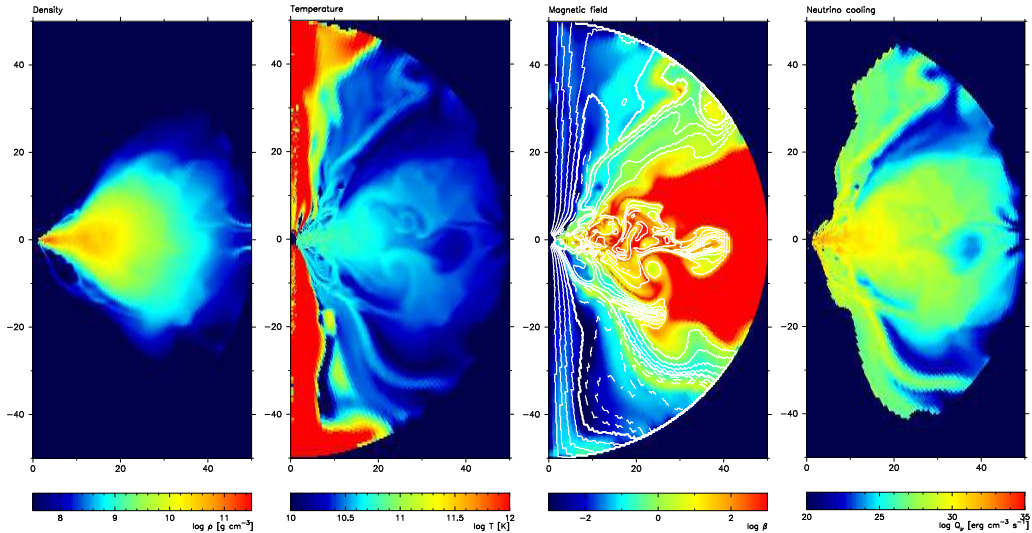

\centering
\includegraphics[width=7cm,angle=270]{Fig3_density_M10_v124.ps}
\includegraphics[width=7cm,angle=270]{Fig3_temperature_M10_v124.ps}
\includegraphics[width=7cm,angle=270]{Fig3_beta_M10_v124.ps}
\includegraphics[width=7cm,angle=270]{Fig3_neu_M10_v124.ps}
\caption{2-D model: Structure of accretion disk in model with neutrino cooling
  taken into account in the dynamical evolution.  The maps show: (i) density,
  (ii) temperature of the plasma, (iii) ratio of gas to magnetic pressure,
  with field lines topology, and (iv) the effective neutrino cooling $Q_{\nu}$
  (from left to right).  Parameters: black hole mass $M = 10M_{\odot}$, spin
  $a=0.98$, initial magnetic field normalization $\beta=50$, and initial disk
  mass $M_{\rm disk}= 1.0 M_{\odot}$. The snapshot is at t = 0.1 s since the
  formation of the black hole.}
\label{fig:torus_bh10_098}
\end{figure*}

In Figures~\ref{fig2} and \ref{fig:torus_bh10_098} we show the maps of the
torus structure calculated in the 2-D model for the black hole mass $M_{\rm BH}=3
M_{\odot}$ and $10 M_{\odot}$, and torus mass of $0.1 M_{\odot}$ and $1.0
M_{\odot}$, respectively (models SBH3 and LBH3 in Table
\ref{table:models}).  The snapshots, taken at the end of the simulation for
time $t=2000 GM_{\rm BH}/c^{3}$, present the baryon density $\rho$, gas
temperature T and magnetic $\beta$ parameter overplotted with magnetic field
lines, as well as the neutrino cooling.

\begin{figure}
\includegraphics[width=7cm]{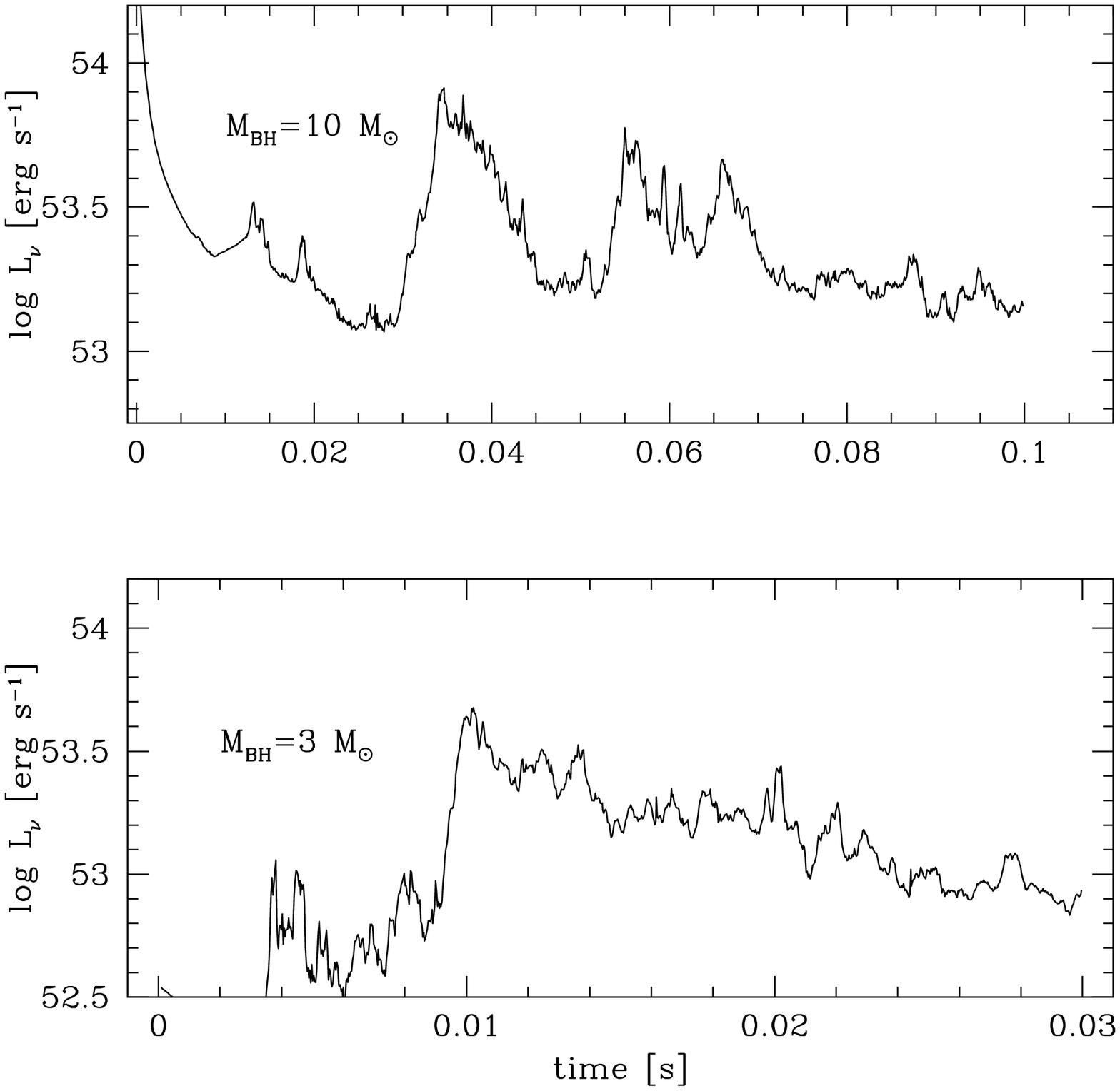}
\caption{Total neutrino luminosity as a function of time.  
The black hole mass is
  $M_{\rm BH}=3 M_{\odot}$ and torus initial mass is 
$M_{\rm d} \sim 0.1 M_{\odot}$ (bottom panel) and $M_{\rm
    BH}=10 M_{\odot}$ and $M_{\rm d} \sim 1.0 M_{\odot}$ (top panel). 
The black hole spin is $a=0.98$.}
\label{fig:neurate_bh3_bh10}
\end{figure}

The neutrino luminosity evolution with time is shown in Figure
\ref{fig:neurate_bh3_bh10} (models with small and large black hole mass).  
For the black hole mass of 3 $M_{\odot}$
and torus of 0.1 $M_{\odot}$, the initial neutrino luminosity calculated using
Eq. \ref{eq:lumin}, is about $10^{52}$ erg s$^{-1}$.  Then the luminosity
gradually grows to over $10^{53}$ erg s$^{-1}$ and peaks at time $t=0.01$ s,
which is equal to about 660 M.  For the black hole mass of 10 $M_{\odot}$ and
more massive torus of 1.0 $M_{\odot}$, the total luminosity is higher and at
maximum reaches values almost $10^{54}$ erg s$^{-1}$, at about $t= 0.04 s$
(equal to about 800 M).  At the end of the simulation, the neutrino luminosity
is about $2\times 10^{53}$ in this model and depends mostly on the ratio
between the torus and black hole mass. The exact values of $L_{\nu}$ at the
end of the simulation are given in Table~\ref{table:models}, 
for a range of parameters.

The neutrinos are emitted from the torus as well as from the hot, rarefied
wind. The luminosity of this wind gives substantial contribution to the total
luminosity and it is about 8-13 \% for SBH models, and 10-15 \% for LBH
models, anticorrelating with the black hole spin.  This fraction was estimated
geometrically, i.e. the wind luminosity was calculated by integrating the
emissivities over the volume above and below $30^{\circ}$ from the mid-plane.
The luminosity of the densest parts of the torus, on the other hand, which can
be estimated e.g. by weighing the total emissivity by the plasma density, is
not more than $10^{48}-10^{49}$ erg s$^{-1}$, because the opacity for
neutrino absorption and scattering in this regions reaches $\tau \sim 0.1$.

\begin{figure}
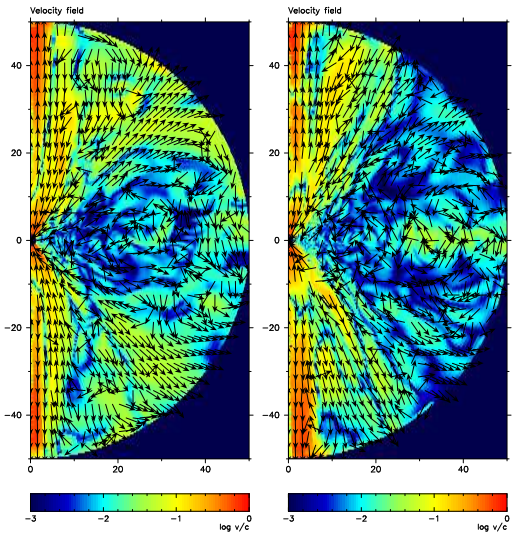

\centering
\includegraphics[width=7cm,angle=270]{Fig6_map.velocity_m3_a0x98_beta100_time0x06.ps}
\includegraphics[width=7cm,angle=270]{Fig6_map.velocity_m10_a0x98_beta100_time0x2.ps}
\caption{Velocity fields at the end of the simulation, $t=4000 M$, 
for black hole mass of
  $3 M_{\odot}$ (left) and $10 M_{\odot}$ (right).  Other parameters: spin $a
  = 0.98$, $\beta=100$. The torus mass is $M_{\rm torus}\approx 0.1 M_{\odot}$
  or 1.0 $M_{\odot}$, respectively.}
\label{fig:velfields}
\end{figure}

The velocity field maps at the end of the simulation, for $M_{\rm BH}=3
M_{\odot}$ and $M_{\rm BH}=10 M_{\odot}$ are shown in
Figure~\ref{fig:velfields}. The figures show results of the models with
highest $\beta=100$ at time t=4000 M, so that we could obtain
clear polar jets.
In the first case, the torus is turbulent, the wind
outflow occurs, but most of material is swept back from the outermost regions
and finally accretes onto black hole. Some fraction of gas is lost via the hot
winds at moderate latitudes. 
In the second model, the disk winds are sweeping the gas out from the system, both in the
equatorial plane and at higher latitudes. 
We identified the regions of the wind in the computation domain 
by defining three conditions that must be satified simultaneously: 
(i) the radial velocity of the plasma is positive 
(ii) the denisty is smaller than $10^{9}$g cm$^{3}$
and (iii) the gas pressure is dominant, $\beta>0.1$. The two latter 
conditions are somewhat arbitrary but they are necessary to
distinguish the wind from the turbulent dense torus and from the
magnetized jets. The winds are located approximately at radii above 10 $R_{\rm g}$ and 
latitudes between about $30^{\circ}-60^{\circ}$ and  $120^{\circ}-150^{\circ}$. 
 The velocity in the wind is 0.005 - 0.18 of the velocity of light (models SBH) and 
0.002 - 0.06 (models LBH). In the first case, it is on the order of the escape velocity, while
in the second case the winds are bound by the black hole gravity (cf., e.g., 
\citet{McKinney06}, who found the winds with half opening angles of 
$\theta=16-45^{\circ}$ and mildly relativistic velocities).
 Such large-scale circulations can be determined in the simulations
with a much larger radial domain (e.g. \citet{Narayan2012, McKinney2012}).

The effect of the wind is the mass loss from the system.  
We estimated quantitatively the
 evolution of the mass during
the simulation. 
The total mass removed from the torus as a function of time,
calculated by integrating the density over the total volume, 
differs significantly from 
the total mass accreted onto the black hole
 (i.e. the time integrated mass accretion rate through the inner boundary,
subtracted from the initial mass).
For models with $M_{\rm BH}=3 M_{\odot}$, 
the denser and cooler torus, with smaller
gas pressure to magnetic pressure ratio, launches a wind and about 50\% of
mass is lost through wind, while the rest is accreted onto black hole.  
However, for the black hole of $10 M_{\odot}$, 
after the wind is launched, it takes
away about 75\% of mass from the system.
In other words, the average mass loss rate in the winds 
is either equal to or larger 
(in particular, in LBH models, it may be even 3 times larger) than the 
accretion rate onto the black hole.
The results are weakly sensitive to the black hole spin value.

The physical conditions in the winds are different from those in the torus. 
The densities are a few orders of magnitude smaller, between 
$5\times 10^{6}$ and $10^{9}$ g cm$^{-3}$, while the temperatures in the wind are very high, in the range $7\times 10^{9} - 5\times 10^{10}$ K (in general, the winds in models LBH are slightly hotter and less dense than in SBH). 
Such high temperatures, above the treshold for electron-positron pair production, $T=m_{\rm e}c^{2}\approx 5 \times 10^{9}$ K, are 
the key condition for neutrino emission processes. The neutrino cooling 
is then efficient and only weakly depends on density. In the clumps with 
$\rho > \sim 10^{8}$ g cm$^{-3}$, the nuclear processes lead to neutrino 
production, while the optical depths for their absorption are very small.

The hot, rarefied, transient polar
jets appear as well on both
sides of the black hole, as seen in Figure \ref{fig:velfields} as well 
as in the maps in Figs. \ref{fig2} and \ref{fig:torus_bh10_098}. 
The limitation of our model is only that 
here we do not study the neutrino emission in these jets.

 In this Section, we show the results of the models where the thickness of the
torus is given by the pressure scale-height at the equator. This is about 0.3
times the radius.  We also tested the approximate condition for the disk
thickness being a fraction of the radius, $H\sim 0.5 r$.
We verified that the disk thickness parametrization of neutrino cooling does not
affect much the accretion rate onto black hole neither the total luminosity.

\subsubsection{Optically thin and thick tori}

\begin{figure}
\includegraphics[width=7cm]{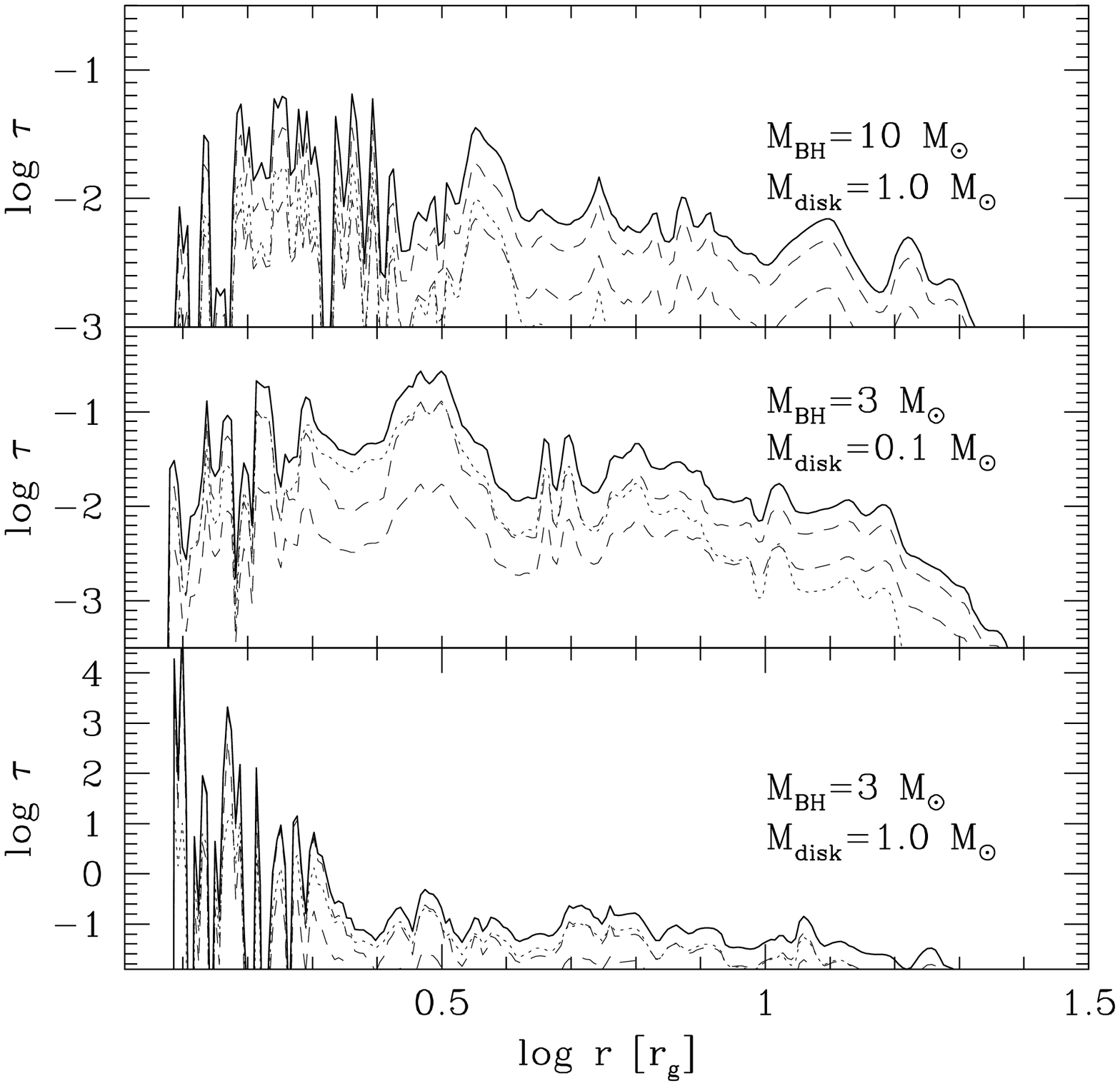}
\caption{Neutrino optical depths due to absorption on tau and muon neutrinos 
 (dashed lines) and
  scattering (dotted lines) and total (solid lines), at the end of the
  simulation for the models with $M_{\rm BH}=3 M_{\odot}$ and torus mass
  $M_{\rm torus} \sim 1.0 M_{\odot}$ (bottom) $M_{\rm BH}=3 M_{\odot}$ and torus mass
  $M_{\rm torus} \sim 0.1 M_{\odot}$ (middle), or $M_{\rm BH}=10 M_{\odot}$
  and $M_{\rm torus} \sim 1.0 M_{\odot}$ (top). The profiles are taken in
  the equatorial plane. The black hole spin is a=0.98.}
\label{fig:optdepth}
\end{figure}

We find no clear neutrinosphere in the models where the
torus to the black hole mass ratio is small and 
the accretion rate is below $\sim 1 M_{\odot}$ s$^{-1}$. In these models,
the torus and wind are
both optically thin to neutrinos and radiate efficiently.  The optical depths
due to the scattering and absorption of neutrinos, calculated in the
equatorial plane, are shown in Figure \ref{fig:optdepth}. As shown 
in the top and middle  panels of the Figure,
$\tau_{\rm tot}\approx 0.15$ in the innermost parts of the torus at the
equator, for the model with black hole mass $M_{\rm BH}=3 M_{\odot}$ and disk
mass of $0.1 M_{\odot}$ (i.e., SBH3 and LBH3). 
Above the equator, the optical depths are much
smaller. Also, the model with back hole mass $M_{\rm BH}=10 M_{\odot}$ and
disk mass of $1.0 M_{\odot}$ gives small neutrino optical depths, up to about
0.05. The flow is optically thin to neutrinos for the
magnetic field parameter $\beta=50 $ as well as $\beta=5$. Therefore the neutrino pressure is
much less than both the gas and magnetic pressures.

\begin{figure}
\includegraphics[width=7cm]{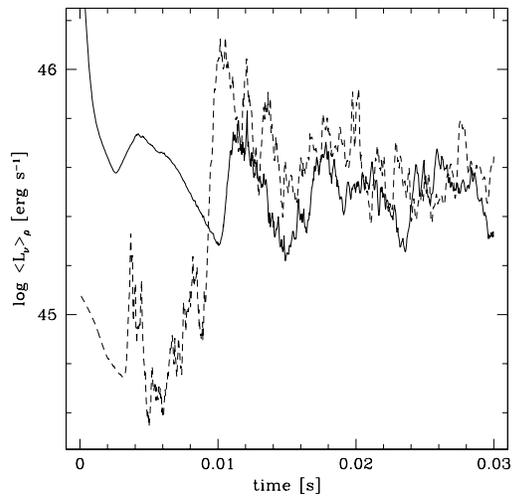}
\caption{Comparison of the optically thin and thick models. 
Neutrino luminosity weighted by the plasma density, at the end of the
  simulation. The models are with  $M_{\rm torus} \sim 1.0 M_{\odot}$ 
  (thick solid line) and  $M_{\rm torus} \sim 0.1 M_{\odot}$ (thin dashed line) 
 The black hole mass is $M_{\rm BH}=3 M_{\odot}$, spin a=0.98 and 
magnetization $\beta=50$.}
\label{fig:lumrho}
\end{figure}

In the bottom panel of the Figure \ref{fig:optdepth}, we show the results 
from the model SBH8, where the torus mass was assumed $1.0 M_{\odot}$ and the black hole mass was
$M_{\rm BH}=3 M_{\odot}$. The accretion rate onto the black hole was in this case larger
than $1.0 M_{\odot}$s$^{-1}$ and the optical thicknesses to the neutrino absorption and scattering
were larger than unity within the inner 3 gravitational radii in the torus equatorial plane.
The neutrino luminosity of the plasma is affected by the opacities. However, the neutrino trapping 
effect that was clearly present in the 1-D models, is now rather subtle and
plays a role in the densest, equatorial regions of the torus.
In Figure \ref{fig:lumrho} we plot the neutrino luminosity
weighted by the plasma density, i.e. $<L_{\nu}>_{\rho} = \int{Q_{\nu}\rho dV}/\int{\rho dV}$.
We see, that after the initial conditions of the simulation are relaxed, about 
0.01 s for the lack hole mass $M_{\rm BH}=3 M_{\odot}$, the luminosity 
of the more massive torus 
drops below the value obtained for the less massive one, optically thin to neutrinos.
Still, the total neutrino luminosity of the system is dominated by the 
optically thin wind, and
the total $L_{\nu}$ of the more massive torus is large (e.g. at $t_{\rm end}$ 
it is equal to $9 \times 10^{52}$ and $4 \times 10^{53}$ erg$s^{-1}$ respectively; 
see Table \ref{table:models}).

In Figure \ref{fig:neurate_bh3_bh10}, we show the 
total neutrino luminosity (i.e. the disk and wind luminosity), in the 
models with different BH and disk mass. These models are optically thin.
In Figure \ref{fig:lumrho}, we show the luminosity weighted by the density,
which represents the densest parts of the disk, where the optical depths 
could be larger than 1.
The meaning of Fig. \ref{fig:lumrho} is therefore 
to compare the optically thick and thin 
models, which have
luminosities slightly different due to neutrino absorption.
Still, the luminosities are on the same order of magnitude, after the initial 
conditions are relaxed.
The differences in the initial conditions leading to the luminosity
differences are mainly due to a larger size and mass of the disk
in the compared
 models, determined by the initial location of the pressure maximum.
After the torus redistributes itself and matter accretes through the
black hole horizon, the initial conditions are relaxed.

\subsubsection{Resolution tests}
\label{sect:resol}

As a standard resolution, we use $256 \times 256$ zones in $r$ and $\theta$.
For numerical test, we also checked two other resolutions, for the model
SBH3. The lowest resolution model was with $128 \times 128$ zones and highest
resolution was with $512 \times 512$ zones.  We found the increase of total
neutrino luminosity with resolution at late times of the evolution, up to a
factor of 2 between the two extreme cases. The time averaged neutrino
luminosity is equal to $4.74 \times 10^{52}$, $1.04 \times 10^{53}$ and $6.55
\times 10^{52}$ erg s$^{-1}$, for the low, medium and high resolution models,
respectively.  Also, the relaxation from initial conditions is reached earlier
for the largest resolution. For the disk structure, the increase of resolution
results in a slight temperature increase and density rise in the inner regions
of the torus, because the magneto-rotational turbulence is better resolved and
accretion rate is increased. The time dependence of accretion rate onto black
hole is finest for highest resolution models. The peaks in the accretion rate
are higher, occur earlier during the evolution and continue to the end of
simulation.

Still, we conclude that it is justified to keep the moderate resolution as the
basic one, as it satisfies the balance between accuracy and computation time.

\subsection{Effects of the black hole spin}

We ran our small and large black hole simulations with three values of the
black hole spin parameters, $a=0.98$, $a=0.9$, and $a=0.8$.  The value of
black hole spin is qualitatively not very significant for the average
properties of the torus.  For the lower spins, the torus is slightly hotter
and less magnetized, with the neutrino emissivity being smaller both in the
torus and in the wind.

The flaring activity, shown in the Figure~\ref{fig:accrate} and discussed
above, is stronger for smaller black hole spins at late times, and the
accretion rate onto black hole occasionally reaches 3-4 or even 5-6
$M_{\odot}$s$^{-1}$, depending on the black hole to torus mass ratio. 
The fast spinning black holes launch powerful and steady polar
jets. However, tha values of the Blandford-Znajek luminosity as given in 
Table \ref{table:models}, do not differ significantly for our spins (a=0.8-0.98).  
 These results should be further verified by the
3-dimensional simulations with a range of grid resolutions.

The mean accretion rate onto the black hole decreases with black hole spin, as
given in Table \ref{table:models}. The result is therefore the same as in
\citet{devillers}, regardless of the neutrino cooling included.

\subsection{Effect of the magnetic field}

The magnetic field in our simulations was parametrized with initial
conditions of $\beta=P_{\rm gas}/P_{\rm mag}$ 
of a fixed value with a maximum at the pressure maximum radius
and zero everywhere outside of the torus.

The mean value of $\beta$, integrated over the total
volume, was at $t=0$ infinite due to such initial conditions, but at the end of the simulation
converged to the value assumed for the torus.  
The mean $\beta$ weighted by the density was always a bit larger than
the total volume integrated beta due to the dominating gas pressure in the disk.

Changing the magnetic field normalization $\beta$ affects somewhat the
resulting structure of the torus.  
The torus
density increases with $\beta$: the maximum density at the equatorial plane
for the torus around a $3 M_{\odot}$ black hole with $\beta_{\rm init}=50$ is
$\rho_{\rm max}\approx 1.5\times10^{12}$ g cm$^{-3}$, for $\beta_{\rm
  init}=10$ it is $3.5\times10^{11}$ g cm$^{-3}$, and for $\beta_{\rm init}=5$
it is $1.5\times10^{11}$ g cm$^{-3}$ (all results are for 
$t=0.03$ s of the
torus evolution; the models we compare are SBH3, SBH4 and SBH5).  Similar
trend in density is found for other torus to black hole mass ratios.  The
temperature of the torus is roughly similar for all the $\beta$ values we
tested and $T_{\rm max}\approx 1.2\times10^{11}$ K, however the jets are cold
only for the highest $\beta$. The latter might be affected by numerical
effects, so we do not analyze the jets structure here.

For the largest $\beta_{\rm init}$ we tested, the contrast between the highly
magnetized polar jets and weakly magnetized disk is most pronounced. For
smaller $\beta$, we have a region of mildly magnetized flow in the
intermediate latitudes.  The speed of evolution of the disk also depends on
$\beta$ and the shortest relaxation time is for the model with smallest
$\beta_{\rm init}$, because the viscous time scale is small in this case.  On
the other hand, the large $\beta$ means that the magnetic field is weak and
therefore the action of magnetic dynamo most quickly dies out.

Also, the accretion rate on average is larger for small $\beta$, i.e.  the
accretion rate correlates with the viscosity, the same as in a standard
accretion disk.  We compared the accretion rates for
several values of $\beta$ parameter. We noticed that 
the flares are higher when $\beta$
decreases, so for the most magnetized plasma we studied, the accretion rate
can reach even 10 $M_{\odot}$s$^{-1}$.

\begin{figure}
\includegraphics[width=7cm]{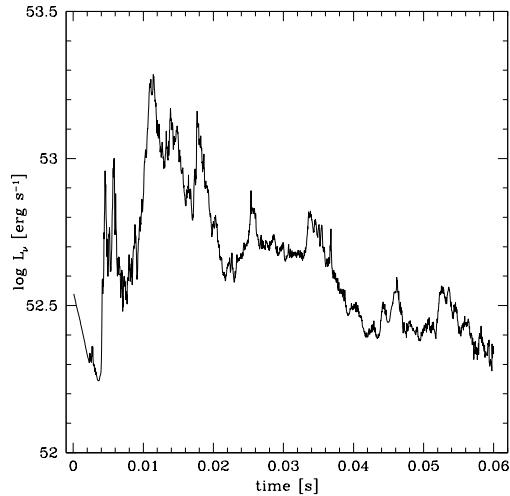}
\caption{Neutrino luminosity as a function of time,
for the neutrino cooled model with black hole mass $3M_{\odot}$ and spin
$a=0.98$, with $\beta=100$.}
\label{fig:lneu_beta}
\end{figure}

In Figure \ref{fig:lneu_beta} we show the neutrino luminosity
for $\beta = 100$.
The general evolution of the luminosity does not depend on $\beta$, 
so the maximal neutrino luminosity is reached at time $\sim 0.01$, 
and then $L_{\nu}$ slowly decreases. The value of the maximum luminosity 
exceeds $2 \times 10^{53}$ erg s$^{-1}$. 
This value does not 
depend significantly on $\beta$ parameter and the differences (see Table \ref{table:models}) should be
attributed mainly to numerical uncertainties (see Section 3.1.2).

The Figure \ref{fig:lneu_beta} shows the simulation up to time 4000 M
(model SBHlb).
for the initial configuration, estimated as the ratio
between the total thermal energy and neutrino luminosity, 
is in this model  equal to 0.12 s, while
in the models SBH4 and SBH5 ($beta=10$ and $\beta=5$), it is 
equal to $\tau_{\nu} \approx 0.05-0.07$ s.

\subsection{Comparison to the models without neutrino cooling}

The torus around the spinning black hole at hyper-Eddington rates is cooled by
neutrinos and in the 1-D simulations the neutrino cooling effects were studied
e.g., by \citet{janiuk}; \citet{chen}.  To quantify the effect of neutrino
cooling in 2D MHD simulations, we ran a test model with no cooling assumed.

\begin{figure}
\includegraphics[width=7cm]{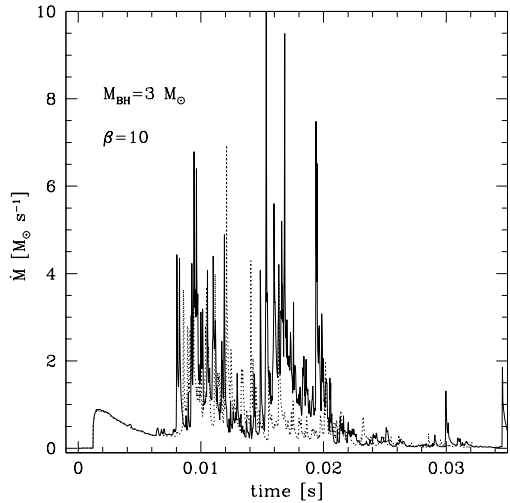}
\caption{Accretion rate as a function of time, in the models with and without neutrino
  cooling (thick solid and thin dashed lines, respectively).  
  The black hole mass is $3M_{\odot}$, its spin is $a=0.98$, and the
  initial disk mass is $0.1 M_{\odot}$.  The initial magnetic field
  normalization is $\beta=10$.}
\label{fig:accrate_beta}
\end{figure}

In Figure \ref{fig:accrate_beta} with a thin dashed line
we plot the accretion rate as a function of
time for an exemplary model without neutrino cooling.  The average accretion rate onto
black hole is lower in these models than in the cooled models, for the same
black hole spin and magnetic field.  Decreasing the $\beta$ parameter,
i.e. increasing the viscosity, results in the increase of the accretion rate,
similarly to the $\alpha$-disks.

The density of the disk in the models without cooling is
smaller in the equatorial plane, the disk being less compact (i.e., less dense
and geometrically thicker) and hotter than in the neutrino-cooled disks.  The
disk without cooling is also more magnetized i.e. the ratio of gas to magnetic
pressure, $\beta$, is on average smaller in the disk.  This is because the
pressure decreases with smaller density, albeit the higher temperatures in the
plasma.

\begin{figure}
\includegraphics[width=7cm]{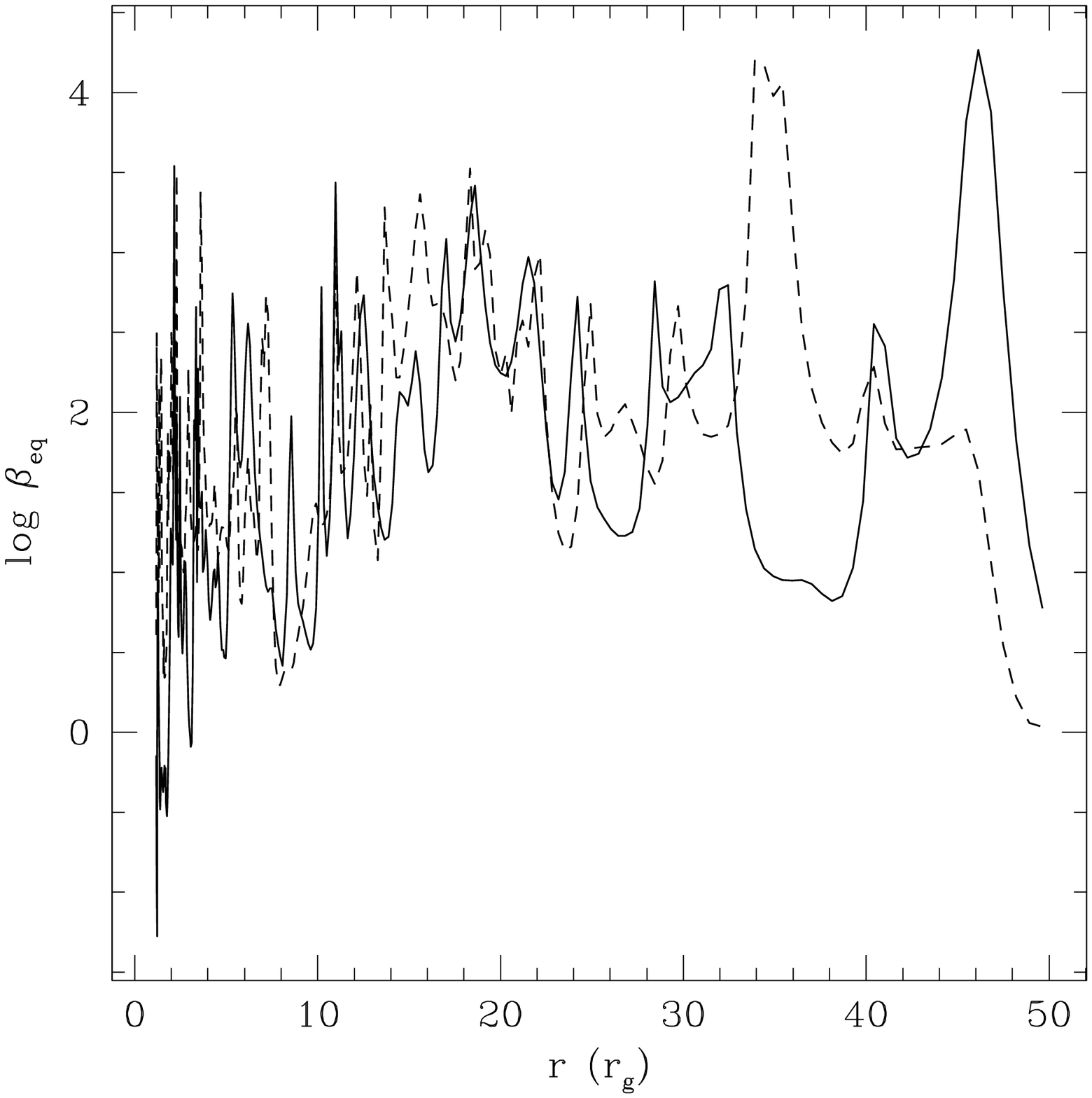}
\caption{The ratio of the gas to magnetic pressure in the equatorial plane of
  the torus in the function of radius, at the end of the simulation ($t_{\rm end}=2000 M$), for the
  models with and without neutrino cooling (solid and dashed lines,
  respectively).  The black hole mass is $3M_{\odot}$, and its spin is a=0.98,
  while the initial disk mass is $0.1 M_{\odot}$, and initial magnetic field
  normalization is $\beta=50$.}
\label{fig:betadisk}
\end{figure}

\begin{figure*}
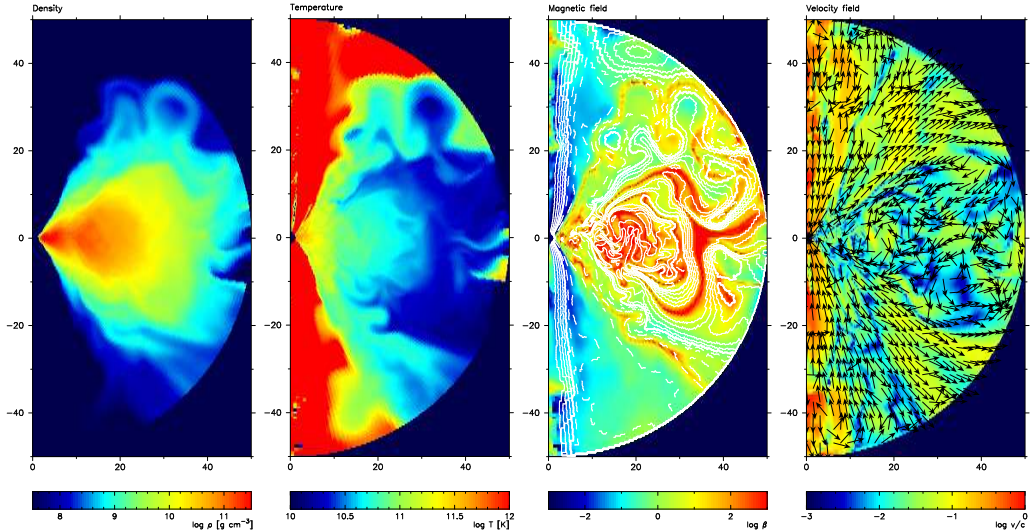

\centering
\includegraphics[width=7cm,angle=270]{Fig13_map.density.bh3.a098.b50.t040.nocool.ps}
\includegraphics[width=7cm,angle=270]{Fig13_map.temperature.bh3.a098.b50.t040.nocool.ps}
\includegraphics[width=7cm,angle=270]{Fig13_map.beta.bh3.a098.b50.t040.nocool.ps}
\includegraphics[width=7cm,angle=270]{Fig13_map.velocity_m3_a0x98_nocool.ps}
\caption{Model without neutrino cooling. The parameters are: $a=0.98$,
  $M_{BH} = 3 M_{\odot}$, $\beta_{\rm init}=50$, $M_{\rm torus}= 0.1
  M_{\odot}$.  The maps, from left to right, show the distribution of density,
  temperature, ratio of gas to magnetic pressure with field lines topology,
  and velocity field.  The snapshot is at t=0.03 s since the formation of the
  black hole.}
\label{fig:torus_nocool_09}
\end{figure*}

The distribution of gas to magnetic pressure in the equatorial plane is shown
in Figure \ref{fig:betadisk}.  The maps of the density, temperature and
magnetic field are shown in Figure \ref{fig:torus_nocool_09}.

\begin{figure}
\includegraphics[width=7cm]{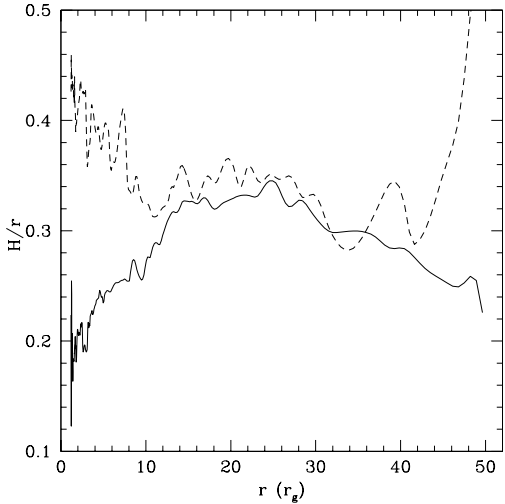}
\caption{The thickness of the torus in the function of radius, at the end of
  the simulation, for the models with and without neutrino cooling (solid and
  dashed lines, respectively). The black hole mass is $3M_{\odot}$, and its
  spin is a=0.98, while the initial disk mass is $0.1 M_{\odot}$, and initial
  magnetic field normalization is $\beta=50$.}
\label{fig:hdisk}
\end{figure}

Also, the thickness of the torus, measured by the pressure scale height 
at the
equator, is larger in case of no neutrino cooling, as shown in the example 
in Figure \ref{fig:hdisk}. 
The ratio of $H/r$ is about 0.3-0.5 in the model
without neutrino cooling, and it is 0.1-0.3 in the cooled disk 
(initial approximation of $H=0.5 r$ was used to compute the neutrino
opacities).

To sum up, the mass accretion rate remains similar, but the structure of the
disk changes, compared to the torus evolving with no neutrino cooling: the
disk is geometrically thinner and more magnetized.

\subsection{Comparison with 1-dimensional models}

In this section, we quantify the effects of 2-dimensional GR MHD approach with
respect to the simplified 1-D neutrino cooled torus model \citep{janiuk} and
compare the 1D and 2D models.  The 1-D model is parametrized by the black hole
mass, spin and $\alpha$ viscosity. To compare its results with the relaxed
model in 2-D simulations, we set these parameters to 3 $M_{\odot}$, 0.98 and
0.1, respectively, which corresponds to the SBH5 2-D model in the Table
\ref{table:models}. The accretion rate is taken equal to $0.17
M_{\odot}$s$^{-1}$ which is the mean acretion rate computed after evolving the
2-D model.

The structure of the disk in our 1-D model is calculated assuming the
zero-torque boundary condition at the marginally stable circular orbit.  Its
location is dependent on the black hole spin, according to the formulae by
\citet{bardeen} (see \citet{janiuk2010} Eq. (17)).  
This condition is used for standard $\alpha$-disks and does
not apply in the MHD simulations.
The total mass of the torus, calculated up to $50
r_{\rm g}$, is computed from integration of the converged
surface density profile.  
The resulting value is of the same order as that assumed  in
the 2-D calculations by defining the location of the pressure maximum, the 
difference being mainly due to lower density in the inner 
$\sim 6 R_{\rm g}$ of the 2-D model equatorial plane.

The viscosity in the 1-D simulations was parametrized by means of the Shakura
\& Sunyaev (1973) $\alpha$ constant. In the 2-D model,
the viscosity is due to the magnetic turbulence, as parametrized with an
initial value of $\beta$ inside the torus and infinite outside it, and
then depending on the location and evolving in time.

The angular momentum is transported outwards due to magneto-rotational
turbulence.  In consequence, no constant value of viscosity is obtained, but
after the initial conditions imposed by $\beta_{\rm init}=P_{\rm gas}/P_{\rm
  mag}$ are relaxed, the system slowly converges to a value $\beta={u
  (\gamma-1) \over 1/2 B^{2}}$, which approximately corresponds to $\alpha$ via
the relation $\alpha \approx 1/(2\beta)$. 
This approximate relation might be verified with a 
3-D model of the magneto-rotational instability with Maxwell 
and Reynolds stresses computed directly.

\begin{figure}
\includegraphics[width=7cm]{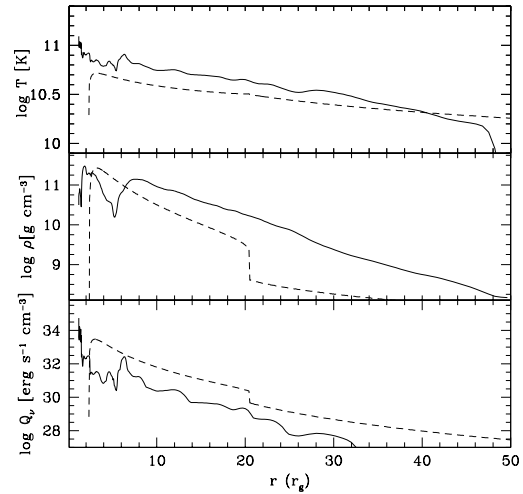}
\caption{Comparison of the 1D model (dashed lines) and 2D GR MHD model (solid lines).
Plots show the  temperature (top panel), density (middle) and 
neutrino cooling rate (bottom panel) in the
  function of radius. Black hole mass is $M_{\rm BH}=3 M_{\odot}$ and its spin is $a=0.98$. 
  The 2-D profiles were taken in the equatorial plane, at the end of the simulation in model 
SBH5l ($t_{\rm end}= 0.15$ s, $\dot M(t_{\rm end}=0.024 M_{\odot}$s$^{-1}$). 
The 1-D profiles are the vertically integrated density and cooling rate,
devided by the pressure scaleheight.
Parameters of the 1-D model are: $t=0$ (i.e. stationary model), $\dot M = 0.024 M_{\odot}$s$^{-1}$, 
viscosity $\alpha=0.1$. }
\label{fig:qnudisk}
\end{figure}

The 2-dimensional structure of the torus is basically consistent with the
results of 1-D models. The results are shown in Figure~\ref{fig:qnudisk}. 
The equatorial 
density profiles have the same average slopes and normalisations are within the 
same order of magnitude, 
up to $20 r_{\rm g}$,
however they differ  due to the
types of boundary conditions.

The temperature profiles have the same slopes in 1-D and 2-D equatorial plane. 
Their relative normalisations differ only slightly and they depend mostly on $\alpha$
 value. 
We note that in the 2-D
models the temperature is more sensitive to resolution, as the MHD turbulence
 is better resolved.

The neutrino cooling profiles in
the 1-D and 2-D models are similar within 2 orders of magnitude. 
At inner parts of the tori the boundary conditions
are different, and at outer parts the neutrino emissivity in 2-D model
decreases due to drop in density and temperature. 
Close to the inner edge of the torus, the emissivity in the 2-D
model strongly varies, because of the magnetic turbulence and thermal
flickering, which was not accounted for in the 1-D model.

\section{Summary and discussion}\label{sec:diss}

\begin{table*}
\begin{center}
\caption[]{Summary of the models. Mass is given in the units of
  $M_{\odot}$, time in seconds and luminosity in erg s$^{-1}$. \label{table:models}
}
\begin{tabular}{ccccccccccc}
\hline
Model & $M_{\rm BH}$ & $a$ & $\beta_{\rm init}$ & $t_{\rm end}$ & $M_{\rm torus}(t=0)$ &  $M_{\rm torus}(t_{\rm end})$ & $<\dot M>$ & $\dot M(t_{\rm end})$ & $L^{\rm tot}_{\nu}(t_{\rm end})$ & $L_{\rm BZ}(t_{\rm end})$ \\ 
\hline   
SBH1 & 3 & 0.8 & 50 & 0.03 & 0.1 & 0.09 & 0.45 & 0.10  & $3.70 \times 10^{52}$ & $3.52 \times 10^{51}$ \\
SBH2 & 3 & 0.9 & 50 & 0.03 & 0.1 & 0.09 & 0.40 & 0.13  & $4.63 \times 10^{52}$ & $9.96 \times 10^{50}$ \\
SBH3 & 3 & 0.98& 50 & 0.03 & 0.1 & 0.10 & 0.31 & 0.14  & $8.89 \times 10^{52}$ & $6.63 \times 10^{50}$ \\
SBH4 & 3 & 0.98 & 10 & 0.03  & 0.1  & 0.06  & 0.86 & 0.22 & $1.25 \times 10^{53}$ & $2.67 \times 10^{51}$ \\
SBH5 & 3 & 0.98 & 5  & 0.03  & 0.1  & 0.05  & 1.01 & 0.66 & $2.93 \times 10^{53}$ & $4.61 \times 10^{51}$ \\

SBH6 & 3 & 0.98 & 50 & 0.03 & 0.8 & 0.65 & 0.78 & 0.42 & $2.45 \times 10^{53}$ & $7.32 \times 10^{51}$ \\
SBH7 & 3 & 0.98 & 50 & 0.03 & 0.3 & 0.27 & 0.50 & 0.31  & $1.88 \times 10^{53}$ & $1.56 \times 10^{51}$ \\
SBH8 & 3 & 0.98 & 50 & 0.03 & 1.0 & 0.87 & 0.96 & 0.41  & $4.06 \times 10^{53}$ & $3.02 \times 10^{51}$ \\


LBH1 & 10& 0.8 & 50 & 0.1  & 1.0   & 0.86 & 0.88 & 0.63   & $1.35\times 10^{53}$ & $1.65 \times 10^{51}$ \\ 
LBH2 & 10& 0.9 & 50 & 0.1  & 1.0  & 0.88  & 0.77 & 0.31   & $1.53\times 10^{53}$ & $3.31 \times 10^{51}$ \\ 
LBH3 & 10& 0.98 & 50 & 0.1  & 1.0 & 0.90  & 0.52 & 0.31    &  $1.92\times 10^{53}$ & $2.27 \times 10^{51}$ \\

LBH4 & 10& 0.98 & 10 & 0.1 & 1.0 & 0.72  & 1.31  & 0.30 & $1.65\times 10^{53}$ & $9.04 \times 10^{51}$  \\ 
LBH5 & 10& 0.98 & 5  & 0.1 & 1.0 & 0.58  & 1.71  & 0.67 & $2.63\times 10^{53}$ & $6.34 \times 10^{51}$  \\ 
LBH6 & 10& 0.98 & 50 & 0.1 & 2.7  & 2.27  & 0.87 & 0.64  & $3.15\times 10^{53}$  & $1.28 \times 10^{52}$ \\ 
LBH7 & 10& 0.98 & 50 & 0.1 & 0.4  & 0.32  & 0.35 & 0.22 & $8.07\times 10^{52}$ & $1.77 \times 10^{51}$  \\ 
\hline
\hline
SBH4l & 3 & 0.98 & 10 & 0.15  & 0.1  & 0.04  & 0.22  & 0.03 & $1.16 \times 10^{52}$ & $1.29 \times 10^{51}$ \\
SBH5l & 3 & 0.98 & 5  & 0.15  & 0.1  & 0.031 & 0.18  & 0.02 & $8.86 \times 10^{51}$ & $7.18 \times 10^{50}$ \\
LBH4l & 10& 0.98 & 10 & 0.5   & 1.0  & 0.59  & 0.34  & 0.04 & $2.43\times 10^{52}$ & $2.44 \times 10^{51}$  \\ 
LBH5l & 10& 0.98 & 5  & 0.5   & 1.0  & 0.47  & 0.42  & 0.05 & $2.74\times 10^{52}$ & $3.89 \times 10^{50}$  \\ 
\hline
\hline
SBHlb & 3 & 0.98 & 100 & 0.06  & 0.1  & 0.03 & 0.11  & 0.06 & $2.16 \times 10^{52}$ & $3.85 \times 10^{50}$ \\
LBHlb & 10& 0.98 & 100 & 0.2   & 1.0  & 0.85  & 0.28  & 0.19 & $6.15 \times 10^{52}$ & $1.22 \times 10^{51}$  \\ 
\end{tabular}
\end{center} 
\end{table*}

We calculated the structure and short-term evolution of a gamma ray burst
central engine in the form of a turbulent torus accreting onto a black hole.
We studied the models with a range of value of the black hole spin, its mass
to the torus mass ratio and magnetization. We found that
(i) in the 2-dimensional computations, the neutrino cooling changes the torus structure,
making it denser, geometrically thinner and less magnetized; (ii) the total
neutrino luminosity reaches $10^{53}-10^{54}$ erg s$^{-1}$, for the torus to
black hole mass ratio 0.03-0.1, and the time of its peak anticorrelates with
the black hole spin; (iii) at the end of the simulation, $t\sim 0.03$ or
$t\sim 0.1$ s for smaller or larger black hole, the neutrino luminosity is
about $10^{52}-10^{53}$ erg s$^{-1}$, increasing with black hole spin; this is
by 1-2 orders of magnitude larger than the Blandford-Znajek luminosity of the
jets computed in our models; (iv) the neutrino cooled torus launches a fast,
rarefied wind that is responsible for a powerful mass outflow, correlated with
the torus to black hole mass ratio; (v) the contribution of the wind to the
total neutrino luminosity is on the order of 10\% and correlates with its
mass; (vi) the density and temperature profiles in the equatorial plane of the
2-dimensional MHD torus are well reproduced by the vertically averaged
profiles calculated in the 1-dimensional $\alpha$-disk model, however in the
latter case the torus is cooler by a factor of 1.5-2; (vii) the neutrino
cooling rates are similar for the inner $\sim 20-30 R_{\rm g}$ in the 1D and
2D calculations.

The structure of the central engine we modeled is relevant for any gamma ray
burst, the free parameters being mainly the black hole spin and initial
magnetic field strength.  Without neutrino cooling, all the results scale with
the black hole mass and the assumed mass and size of the initial torus.  Here
we have shown only the short timescale calculations, with no extra inflow of
matter to the outer edge of the disk, which would be relevant for the subclass of
long GRBs central engines.  The internal structure of the torus should not
depend on that, as supported e.g. by the recent observations by {\it
  Swift} showing that flares in both short and long GRBs are likely produced
by the same intrinsic mechanism \citep{Margutti}.  In the short GRB models,
during the evolution of the post-merger disks the rings of material of a mass
between 0.01 and 0.1 $M_{\odot}$ can fall back from the eccentric orbits. In
this way, the neutrino luminosity may brighten a few times on a timescale of
$>1$ second \citep{Lee09}.  Mass fallback from the stellar envelope material
is also a key feature of the collapsar model for the long GRBs.

The mass of the torus assumed in most of our models is about 0.1-1.0
$M_{\odot}$, when the black hole mass is fixed at 3 or 10 $M_{\odot}$. A more
massive torus, which can form in the center of a massive star as a 'collapsar'
central engine, would result in accreting a substantial amount of mass and
angular momentum onto the black hole. Therefore the evolution of the black
hole mass and spin should consistently be taken into account, as shown e.g. by
\citep{moderski}.  This is currently neglected in our calculations, and we
focus on the torus much less massive than the accreting black hole, $M_{\rm
  torus}/M_{\rm BH} \le 0.25$. This is still relevant for
 the compact binary merger scenario.

The initial conditions used in our models, similarly to other simulations,
is based on the equlibrium torus solution and embedded magnetic field 
of a specified topology and stregnth.
The recently simulated mergers of hypermassive neutron stars
(e.g. \citet{Shibata2011})  follow the evolution of matter and electromagnetic 
energy ejection during several tens of milliseconds and show that already
at this stage the toroidal magnetic field component is developed and 
relativistic outflows occur. 
Then, it is expected that the neutron star will eventually collapse to
a black hole, after a substantial loss of the angular momentum
due to the gravitational wave emission, and the transient torus 
with a lifetime of about 100 milliseconds will power the GRB engine.
Our simulation covers this last stage of the event; obviously conditions 
for initial magnetic field are mostly artificial at t=0. However, the toroidal
field forms in our computations really quickly, i.e. after one orbit, 
and the evolution of the neutrino luminosity and flares should match 
then the outcome of the former compact object merger.
The  black hole-neutron stars merger simulations
(for a review see \citet{ShibataTaniguchi2011})  
lead mostly to the formation of a massive black hole with a 
remnant disk of less than 10 \% of the total inital mass of the binary. 
Its density depends on the initial mass ratio and primary BH spin, as well 
as on the neutron star's EOS. The final BH spin is determined mostly 
by its initial value. Overall, the coalescence of high mass 
ratio binaries with $a\le0.75$ is a promissing channel for a short GRB progenitor, 
forming a massive disk plus BH system. 
Our simulations are aimed to realize this scenario.
More detailed studies of the dynamical evolution
of the post-merger system, with initial conditions based on
 the direct output of the merger simulations rather than the
 quasi-steady-state torus,
are planned for our future work (see e.g. by \citet{Schwab2012} for
 the post-merger evolution of binary white dwarfs).

The distribution of the compact binaries from the population synthesis models
shows two peaks: double black holes constitute about two-thirds of the
population, while the double neutron star binaries are about 28\%
\citep{belczyn}.  The remaining pairs can contain a low mass black hole and a
neutron star system.  However, as recently computed by Dominik et al. (2011;
in preparation), the most compact binary pairs contain a neutron star and a
black hole of mass 7-13 $M_{\odot}$. The details of the mass distribution
depend on the evolutionary scenario (presence of the common envelope phase)
and are sensitive to the assumed metallicity.  Therefore, a plausible short
GRB scenario may involve a 3 $M_{\odot}$ black hole with a small disk, as well
as a black hole of $M_{\rm BH}= 10 M_{\odot}$.

The luminosity of the torus is comparable to that obtained from relativistic
hydrodynamical simulations \citep{Jaroszynski1993, Jaroszynski1996,
  Birkl2007}.  Also, the relativistic MHD simulations by \citet{Shibata2007}
reported the neutrino luminosity on the order of $L_{\nu} \sim 10^{54}$ erg
s$^{-1}$, depending on black hole spin ($a \le 0.9$) and torus mass. To
compute the electromagnetic luminosity of the observed GRBs, one needs to
consider the efficiency of neutrino-antineutrino annihilation process, as well
as swallowing of some fraction of neutrinos by the black hole due to the
curvature effects. Most of the neutrinos are formed within $10 R_{\rm g}$. The
luminosity obtained in our simulation will lead to the annihilation luminosity
on the order of $L_{\nu \bar\nu} \approx$ a few times $10^{50}$ erg s$^{-1}$
\citep{zalamea2011}, providing an additional energy reservoir to power the GRB
jet. This is on the same order of magnitude as the Blandford-Znajek 
luminosity in the polar jets. The jet power
can be calculated from our models by integrating the 
electromagnetic energy flux on the black hole horizon over the surface area 
\citep{McKinneyGammie2004}. Depending on black hole spin it reaches the values 
in the range of $L_{\rm BZ} \sim 4\times 10^{50} - 3\times 10^{52}$ erg s$^{-1}$, 
consistently with other estimates \citep{Lee2000, KomisBar2009}. 
For the same black 
hole spin and magnetic $\beta$ parameter, the
models with neutrino cooling give about a factor of two smaller $L_{\rm BZ}$ 
than the non-cooled models.

Our results show that the disks around larger mass black holes are in general
less dense and cooler, for the same black hole spin and accretion rate. They
are however brighter in neutrinos, as their peak luminosity scales directly
with mass.  The wind outflows launched form the surface of the accreting torus
are driven by magnetic pressure which can also halt the accretion rate onto
black hole.  The wind is bright in neutrinos, giving an additional
contribution to the total luminosity of the system.

The general relativistic simulations that ignore the radiative (and neutrino)
cooling have recently been discussed e.g. in ref \citet{McKinney2012}.
They discuss various topologies and stregths of initial magnetic field
and  confirm that the value of initial $\beta$ parameter affects
the final, or time-averaged, viscosity. The latter might be to some extent
verified by the observations of accreting X-ray sources (see \citet{King07}),
to help determine on whether the $\alpha$ scales with only magnetic or
the total pressure.
We note that in our simulations the limitations of assumed axisymmetry in 
the model do not allow to fully constrain effective $\alpha$.

The simulations presented in 
\citet{Krolik2005} 
show the existence of the polar jet outflows. The authors do not discuss
massive winds, as they concentrate mostly on the
accretion disk properties. However, \citet{McKinney06} reports on the existence of
winds with moderately relativistic velocities ($\Gamma \sim 1.5$) and
half opening angles of 16-45$^{\circ}$.

The results shown in this work are obtained with a detailed neutrino cooling
description in which we have incorporated the chemical composition of nuclear
matter where the reactions lead to the neutrino production \citep{janiuk}.
The simulations discussed in \citet{dibi2012} 
 include the radiative cooling 
for low luminosities and accretion rates, appropriate for the case of
radiativily inefficient flows in AGN. The scale height of the disk 
in their results is affected by the radiative 
cooling by a factor of 30-50 per cent,
however the density and thickness of the
inner torus might still be partly affected by the initial conditions
assumed in these simulations.
Qualitatively, our results are similar to theirs, as the
neutrino cooling also leads to the denser and thinner
torus inside 10-15 gravitational radii.
The 'bump' outside that radius, seen in the final snapshots 
from our simulations, may partly also be affected by initial
conditions. However, the difference may also arise because of
a stronger radial dependence of neutrino cooling than it is
in the case of photon cooling.
Similarly to \citet{dibi2012}, 
our dynamical model uses a simplified version of EOS. We note that
 the electrons are degenerate near the disk equatorial plane between the BH horizon
and $r\approx 20 R_{\rm g}$, e.g. in the model SBH2. In this small region,
the dynamical computations with $\gamma=4/3$ might not be suitable to describe
the degenerate electrons (see \citet{bkomis08, bkomis10}).  To model
degenerate gas one could introduce a new equation of state
(e.g. $P=P(\rho_{0}) ~ \rho_{0} ^ 1/n$ where $\rho_{0}$ is the density of the
electrons and $n$ is a politropic index (see Paschalidis et al. 2011; Malone
et al. 1975). The latter however is a mayor change of the numerical scheme
since the matter is composed of also partially degenerate and non-degenerate
electrons, protons, helium nuclei and neutrons which can still be described by
perfect gas law.  Moreover, to account for the pressure of photons and
neutrinos one would need to follow the evolution of radiation and neutrino
energy-momentum tensor coupled to the evolution of matter.  Sill, in our
present model the energy carried out from the system by the neutrinos does not
depend on the EOS used in the interior of the disk and most of the energy is
generated in the disk wind. Of course, it is possible that the change of the
EOS would influence the wind strength, structure and neutrino luminosity.  It
would be interesting to explore the wind launching mechanism in this case and
we plan to study this in future work.

\section*{Acknowledgments}

We thank Chris Belczynski, Michal Dominik, Bozena Czerny and Marek Sikora 
for helpful discussions. We also thank the anonymous referee for 
insightful comments.
This research was supported in part by grant NN 203 512638 
from the Polish Ministry of Science and Higher Education.

\end{document}